\documentclass[11pt]{article}
\usepackage{moriond,epsfig}

\def\Journal#1#2#3#4{{#1} {\bf #2}, #3 (#4)}

\def\NPB{{\em Nucl. Phys.} B}

\def\PRD{{\em Phys. Rev.} D}

\def\be{\begin{equation}}
\def\ee{\end{equation}}
\def\bea{\begin{eqnarray}}
\def\eea{\end{eqnarray}}

\newcommand{\metb}{$E_{\rm{T}}\hspace{-1.1em}/$\hspace{1.0em}}
\newcommand{\metc}{E_{\rm{T}}\hspace{-1.1em}/\hspace{0.6em}}

\newcommand{\sstop} {\tilde{t}}
\newcommand{\ssbottom} {\tilde{b}}
\newcommand{\schi} {\tilde{\chi}}

\newcommand{\slep} {\tilde{l}}
\newcommand{\sg}   {\tilde{g}}
\newcommand{\sq}   {\tilde{q}}
\newcommand{\sG}   {\tilde{G}}

\begin{document}
\vspace*{4cm}
\title{SEARCH FOR SUPERSYMMETRY AT THE TEVATRON}

\author{SONG MING WANG\\
(For the CDF and D\O\ Collaborations)}

\address{Institute of Physics, Academia Sinica\\
Taipei, Taiwan 11529, Republic of China}

\maketitle\abstracts{
This paper reviews some of the most recent results from CDF and D\O\ experiments
on searches for supersymmetry (SUSY) at the Tevatron. We focus on searches for
chargino/neutralino, stop, sbottom, and long lived massive SUSY
particles, on data samples up to $\sim 1$ fb$^{-1}$. No signal was observed,
and constraints are set on the SUSY parameter space.
}

\section{Introduction}
Supersymmetry is an extension to the standard model (SM) of particle physics
that overcomes some of the theoretical problem in the SM by introducing a new
symmetry between fermions and bosons. In this model each SM particle has
a superpartner with spin differ by $1/2$.
The symmetry in SUSY is believed to be broken, and several models
(mSUGRA, GMSB, AMSB,...) have been developed to describe the breaking
mechanism.
In some SUSY models the $R-{\rm{parity}}$
\footnote{$R-{\rm{parity}}=(-1)^{3(B-L)+2s}$} quantum number is conserved.
In this case SUSY particles must be produced in pairs, and the lightest
SUSY particle (LSP) is stable. Cosmological constraints require the
LSP to be neutral and colorless \cite{SUSYLSP}.
Thus, the LSP interacts only weakly and escapes detection leading to
experimental signature of missing transverse energy (\metb).
In the mSUGRA model, the LSP is usually the lightest neutralino ($\schi^{0}_{1}$),
and in the GMSB model, the LSP is the gravitino ($\sG$).
CDF and D\O\ search for SUSY based on their unique signatures
in the final states. In this paper we report on recent SUSY search results
from CDF and D\O\ based on $\sim 300$ pb$^{-1}$
to $\sim 1$ fb$^{-1}$ of $p\bar{p}$ collision data at $\sqrt{s}=1.96$ TeV.
The limits presented in this paper are all obtained at 95\% confidence level
(C.L.) .

\section{Searches for Chargino and Neutralino}
\begin{figure}[htbp]
\includegraphics[width=0.32\textwidth]{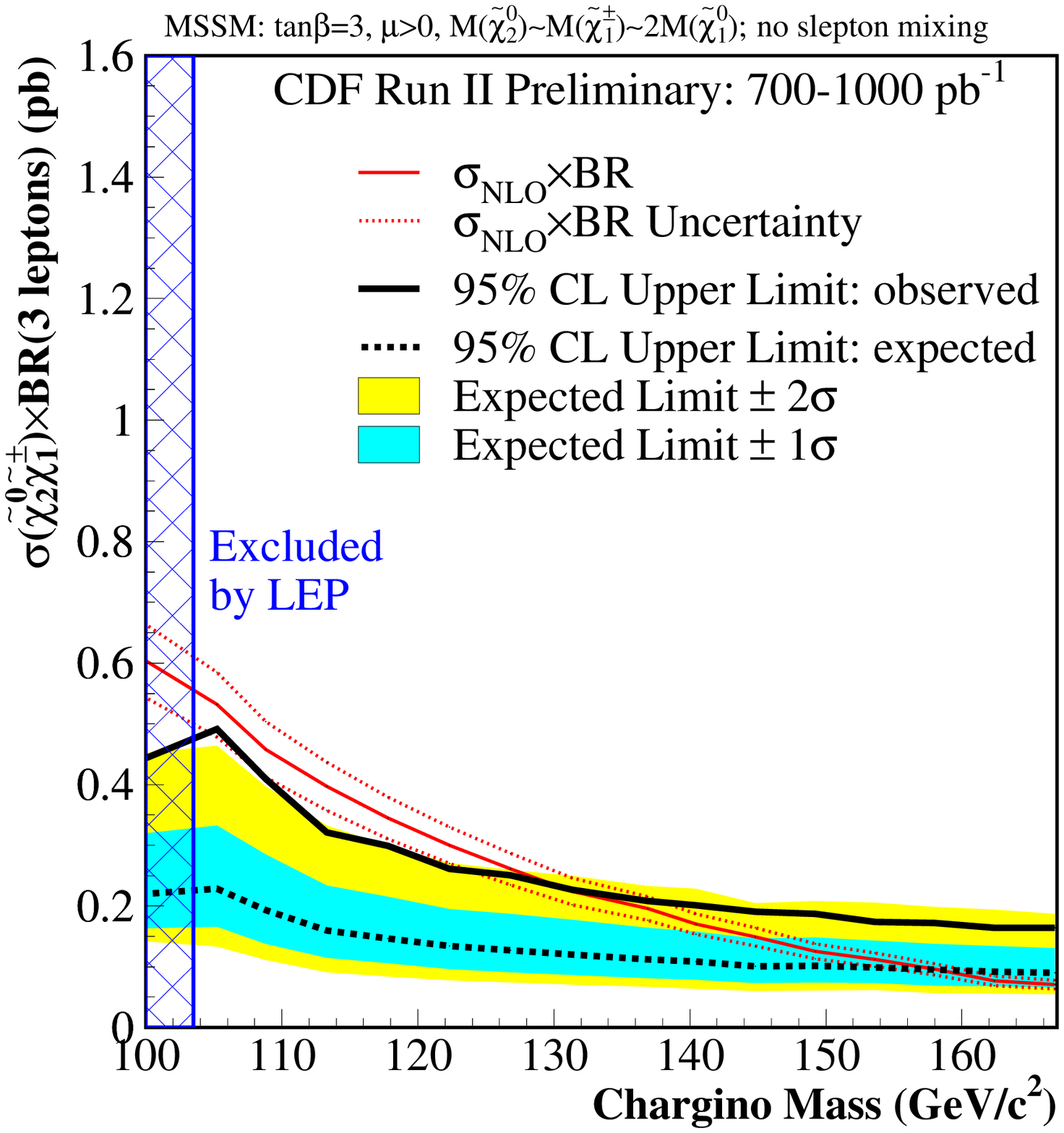}
\hfill
\includegraphics[width=0.32\textwidth]{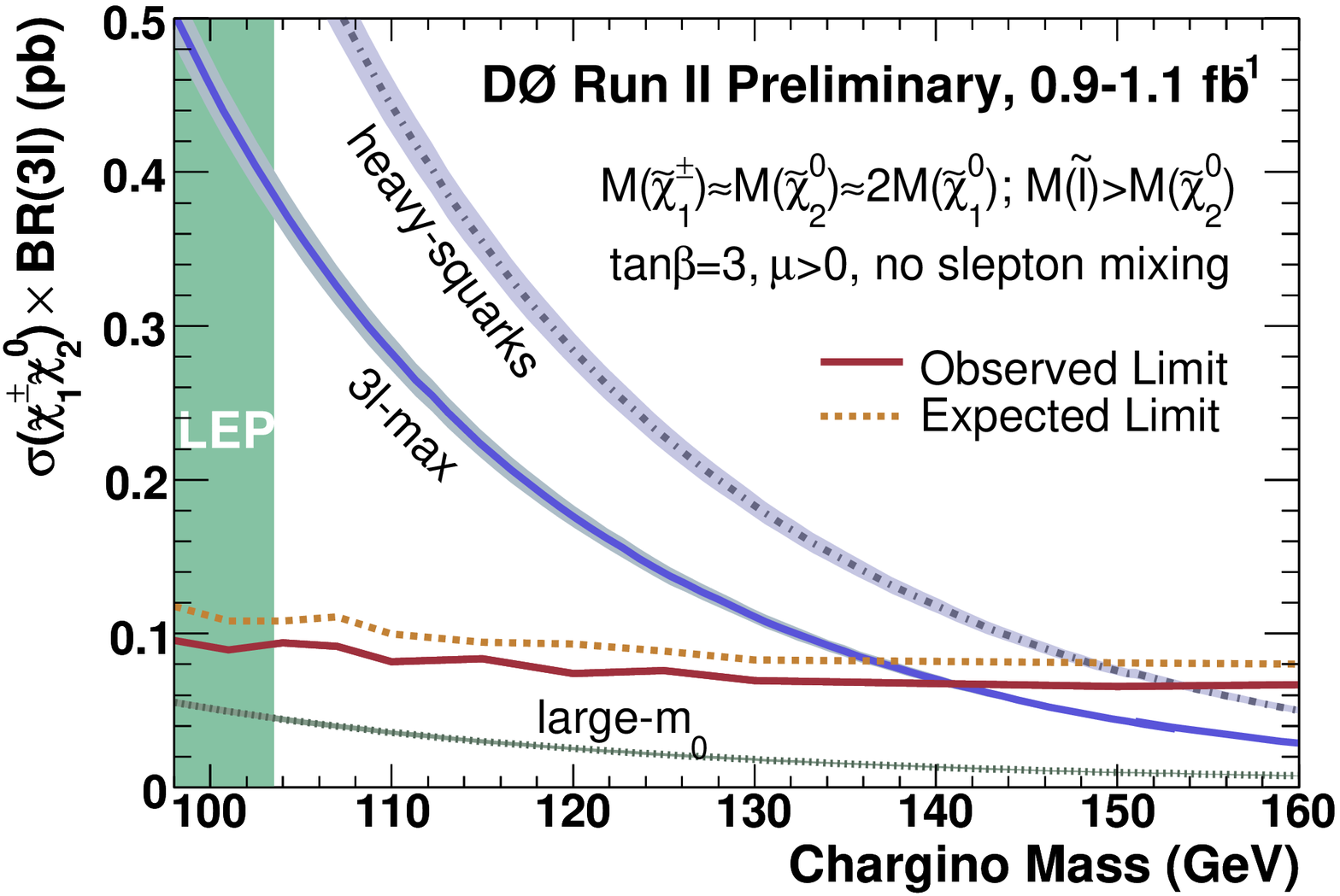}
\hfill
\includegraphics[width=0.32\textwidth]{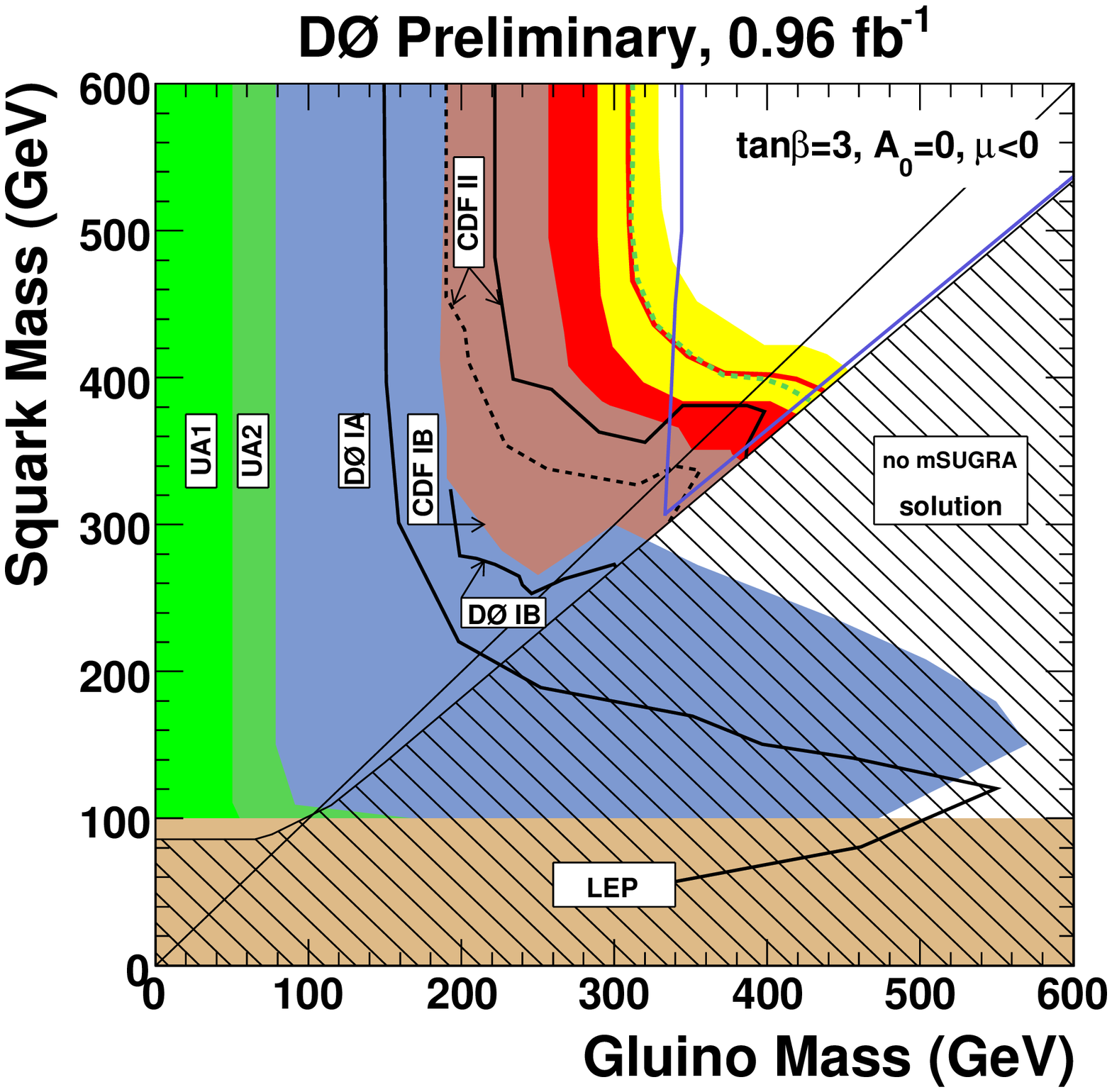}
\\
\parbox[t]{0.32\textwidth}{\caption{The 95\% C.L. upper limit on $\sigma \times BR$
for $\schi^{\pm}_{1}$ and $\schi^{0}_{2}$ production as function of $\schi^{\pm}_{1}$ mass from CDF.}
\label{fig:cdf_C1N2_limit}}
\hfill
\parbox[t]{0.32\textwidth}{\caption{The 95\% C.L. upper limit on $\sigma \times BR$
for $\schi^{\pm}_{1}$ and $\schi^{0}_{2}$ production as function of $\schi^{\pm}_{1}$ mass from D\O\ .}
\label{fig:d0_C1N2_limit}}
\hfill
\parbox[t]{0.32\textwidth}{\caption{The 95\% C.L. excluded region for the squark and gluino
                                    pair production search by D\O\ .}
\label{fig:squarkgluino}}

\end{figure}

The cross section for associated production of chargino ($\schi^{\pm}_{1}$)
and neutralino ($\schi^{0}_{2}$) is expected to be relatively small compared
to other SUSY production via strong interactions.
However $\schi^{\pm}_{1}$ and $\schi^{0}_{2}$ can decay leptonically
($\schi^{\pm}_{1} \rightarrow l \nu \schi^{0}_{1}$,
 $\schi^{0}_{2} \rightarrow l^{\pm} l^{\mp} \schi^{0}_{1}$)
leading to final states with multiple leptons and large \metb from
the escaping $\nu$ and $\schi^{0}_{1}$.
This is a very clean final state since the SM background contribution
is very small. The background is mainly from production of di-boson,
$Z/\gamma^{*}+$jets, $W+$jets, and $t\bar{t}$.
CDF and D\O\ search for $\schi^{\pm}_{1}$ and $\schi^{0}_{2}$ production
in 3-lepton plus \metb final states, and in two like-sign lepton
plus \metb final states, in a data sample of $\sim 1$ fb$^{-1}$.
The number of observed data events and expected SM background events,
after applying all the selection cuts, are shown in Table \ref{tab:C1N2Evt}.
The result indicate no evidence of SUSY
and both experiments extract the 95\% C.L. upper limit
production cross sections, which are shown in
Figure \ref{fig:cdf_C1N2_limit} and \ref{fig:d0_C1N2_limit}.
The limits are obtained in the mSUGRA framework, with $\tan\beta=3$,
and with no slepton mixing.
The CDF's result exclude chargino mass below 130 GeV/c$^{2}$.
The D\O\ \hspace{-0.4em}{'s} result exclude chargino mass below 141 GeV/c$^{2}$,
in the scenario where $m(\slep) \sim m(\schi^{0}_{2})$ such that
the leptonic decay is maximally enhanced.

\begin{table}[htbp]
  \caption{The number of observed data events and expected SM background events
           in the search for associated production of chargino and neutralino.
           CDF's $ee+l$ and $\mu\mu + l$ channels are performed on samples
           collected with low $P_{\rm T}$ threshold.
	   The D\O\ \hspace{-0.4em}{'s} like-sign channel only consists of
           $\mu^{\pm}\mu^{\pm}$.} 
  \label{tab:C1N2Evt}
  \begin{center}
    \begin{tabular}{|l|l|c|c|c|c|c|c|}  \hline
      & Channel &  $ee+l$ & $\mu\mu + l$ & $ell$ & $\mu ll$ & $e\mu l$ & $l^{\pm}l^{\pm}$ \\ \hline
      & $L$ (fb$^{-1}$) & 1 & 1 & 1 & 0.75 &  & 1 \\ \cline{2-8}
CDF   & \# Observed & 3 & 1 & 0 & 1 &   & 13 \\ \cline{2-8}
      & \# SM Expected & $0.97 \pm 0.28$ & $0.40 \pm 0.12$ & $0.75 \pm 0.36$ & $1.26 \pm 0.27$ &   & $7.8 \pm 1.1$ \\ \hline
\hline
      & $L$ (fb$^{-1}$) & 1.1 & 1.1 &   &   & 1.1 & 0.9 \\ \cline{2-8}
D\O\  & \# Observed & 0 & 2 &   &   & 0  & 1 \\ \cline{2-8}
      & \# SM Expected & $0.76 \pm 0.67$ & $0.32^{+0.73}_{-0.03}$ &   &   & $0.94^{+0.40}_{-0.13}$ & $1.1 \pm 0.4$  \\ \hline
    \end{tabular}
  \end{center}
\end{table}

\section{Searches for SUSY in Jets and Missing Energy}

\subsection{Squark and Gluino}
At the Tevatron squark ($\sq$) gluino ($\sg$) can be pair produced, and their
cascade decays can lead to multiple jets and large \metb final state.
The main SM backgrounds are from the production of $Z(\rightarrow \nu \nu)+$jets,
$W(\rightarrow l \nu)+$jets, $t\bar{t}$, and QCD multijets. The \metb
in QCD multijet background is usually a result of jet energy mis-measurement.
D\O\ search for pair production of squark and gluino in the 2, 3 and 4-jet
plus \metb final states, which are respectively, optimized for
$\sq\sq$, $\sq\sg$, and $\sg\sg$ productions.
To select these events in a data sample of 0.96 fb$^{-1}$, cuts on the jets'
transverse energy ($E_{\rm T}$), the event $H_{\rm T}=\sum_{\rm jets} E_{\rm T}$,
and \metb are applied. 
No evidence of squark or gluino production is found, and a limit
in the mSUGRA framework is set as a function of squark and gluino mass.
This is shown in Figure \ref{fig:squarkgluino}.
CDF has also performed a similar search in the 3-jet plus \metb final state,
using a data sample of 1.1 fb$^{-1}$.
Its selection cuts are optimized for different gluino mass ranges.
Similarly no evidence of SUSY is observed, and CDF excludes glunino mass
up to $m(\sg)>380$ GeV/c$^{2}$ for the case where $m(\sg) \sim m(\sq)$.

\subsection{Stop and Sbottom}
In SUSY, one of the stop and sbottom squarks can be much lighter than
the other light flavor squarks. This is due to possible large mixing in the
left and right handed weak eigenstates. Therefore, stop and sbottom may
be produced at a relatively higher rate than the other squarks at the
Tevatron.
CDF and D\O\ search for pair production of stop and sbottom in the
heavy flavor jets and \metb final state, using data samples of
$\sim 300$ pb$^{-1}$.
In these searches it is assumed that the stop decay via a flavor changing
neutral current loop ($\sstop_{1} \rightarrow c \schi^{0}_{1}$),
and the sbottom decay in the channel $\ssbottom_{1} \rightarrow b \schi^{0}_{1}$.
To select these events, both experiments require two or three jets in the
final state, at least one of the jet tagged as a $c$ or $b$ jet, and the
event should have large \metb ($\metc > 50$ GeV) and no isolated lepton.
After the event selections, the observed data events are consistent with
the expected SM backgrounds, which are mainly from $Z(\rightarrow \nu \nu)+$jets,
$W(\rightarrow l \nu)+$jets, $t\bar{t}$, and QCD multijets.
The interpretation of the null results from both stop and sbottom searches
are presented as a 95\% C.L. exclusion region in the mass planes of
$m(\schi^{0}_{1})$ vs $m(\sstop_{1})$ (Figure \ref{fig:stop}),
and $m(\schi^{0}_{1})$ vs $m(\ssbottom_{1})$.
Both CDF and D\O\ exclude stop mass up to $\sim 140$ GeV/c$^{2}$
at $m(\schi^{0}_{1})=55$ GeV/c$^{2}$.
The sbottom mass is excluded up to 222 GeV/c$^{2}$ by D\O\ ,
and 195 GeV/c$^{2}$ by CDF.

\section{Searches for Long Lived SUSY Particles}

\subsection{Searches for Long Lived Neutrino}
CDF has performed a search for massive long lived particles
that decay into photons inside the detector.
The search is focused on the GMSB model where the neutralino
$\schi^{0}_{1}$, which is the next lightest SUSY particle (NLSP),
is long lived and decays into a $\gamma$ and a $\sG$ inside the detector.
The $\sG$ escapes detection and this leads to \metb signature
in the final state.
The $\gamma$ from the $\schi^{0}_{1}$ decay arrives at the face
of the detector at a later time compared to the other photons
that are promptly produced at the interaction point.
For this search the events are selected with at least
one photon with $E_{\rm T}>30$ GeV, one or more jet with
$E_{\rm T}>35$ GeV, and $\metc>40$ GeV.
The arrival time of the photon candidate is measured in the
electromagnetic section of the calorimeter, and is corrected
for the time-of-flight assuming that the photon is coming
from the primary interaction point.
The background consists of prompt photons produced in $p\bar{p}$ collisions,
and fake photons from beam halo and cosmic.
The signal is searched in the photon time window of $2-10$ ns.
In this time window two events are observed in the data,
and $1.3\pm0.7$ events are expected from the background.
Using this result, an exclusion region in the neutralino lifetime
vs neutralino mass plane is determined, and it is shown in
Figure \ref{fig:delayphoton}.

\subsection{Searches for Charge Massive Particles}
\begin{figure}[htbp]
\includegraphics[width=0.32\textwidth]{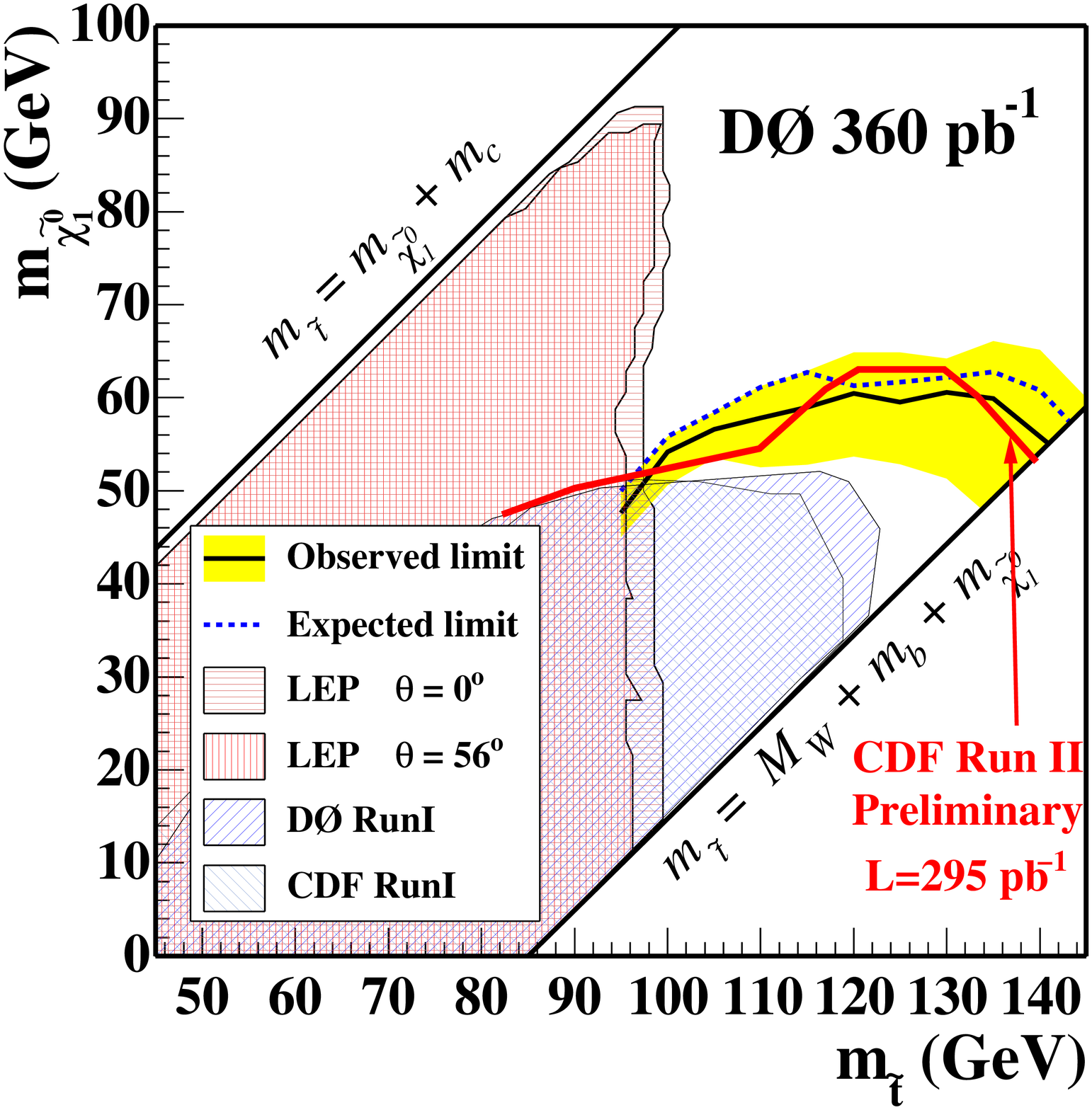}
\hfill
\includegraphics[width=0.32\textwidth]{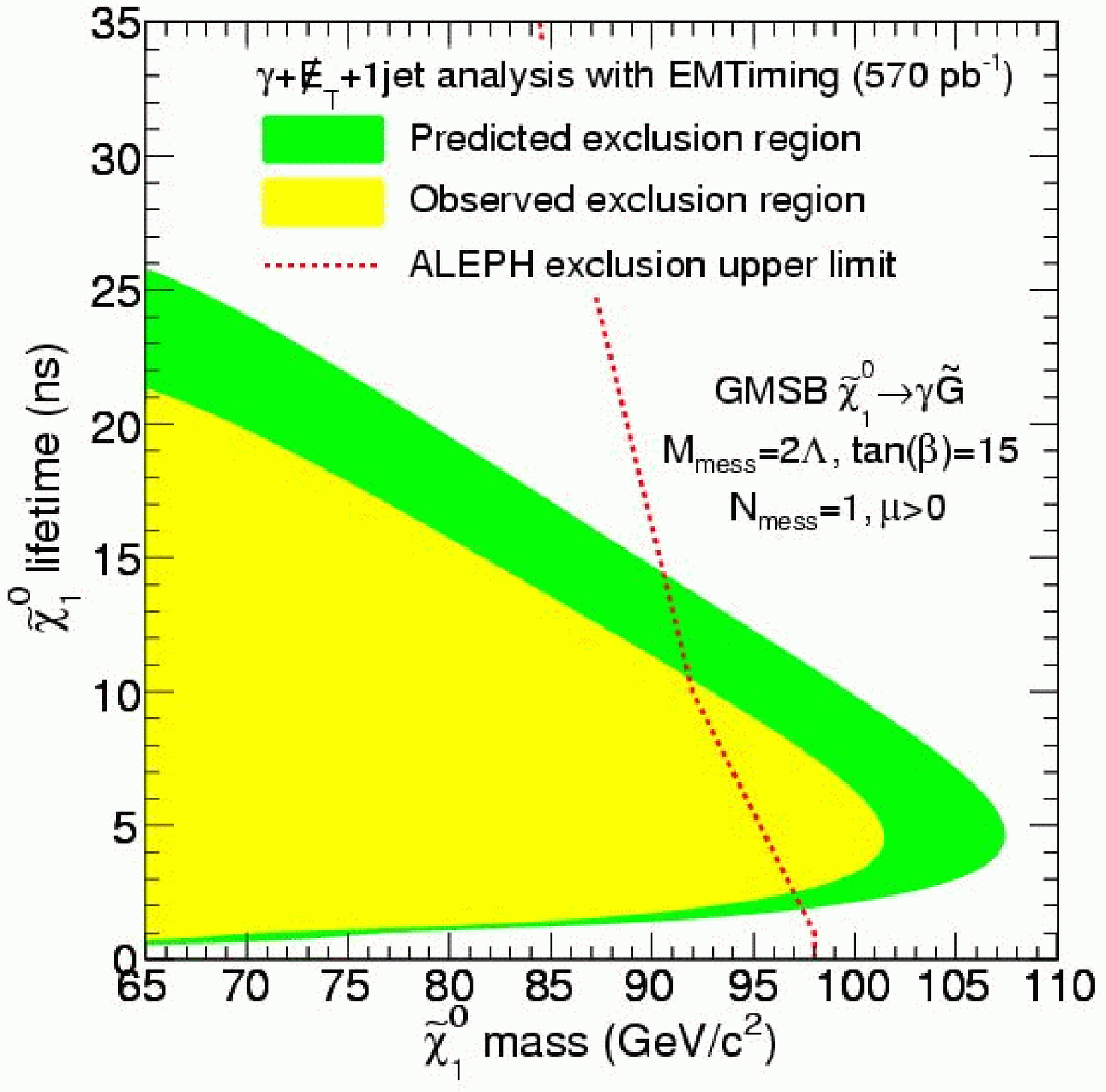}
\hfill
\includegraphics[width=0.32\textwidth]{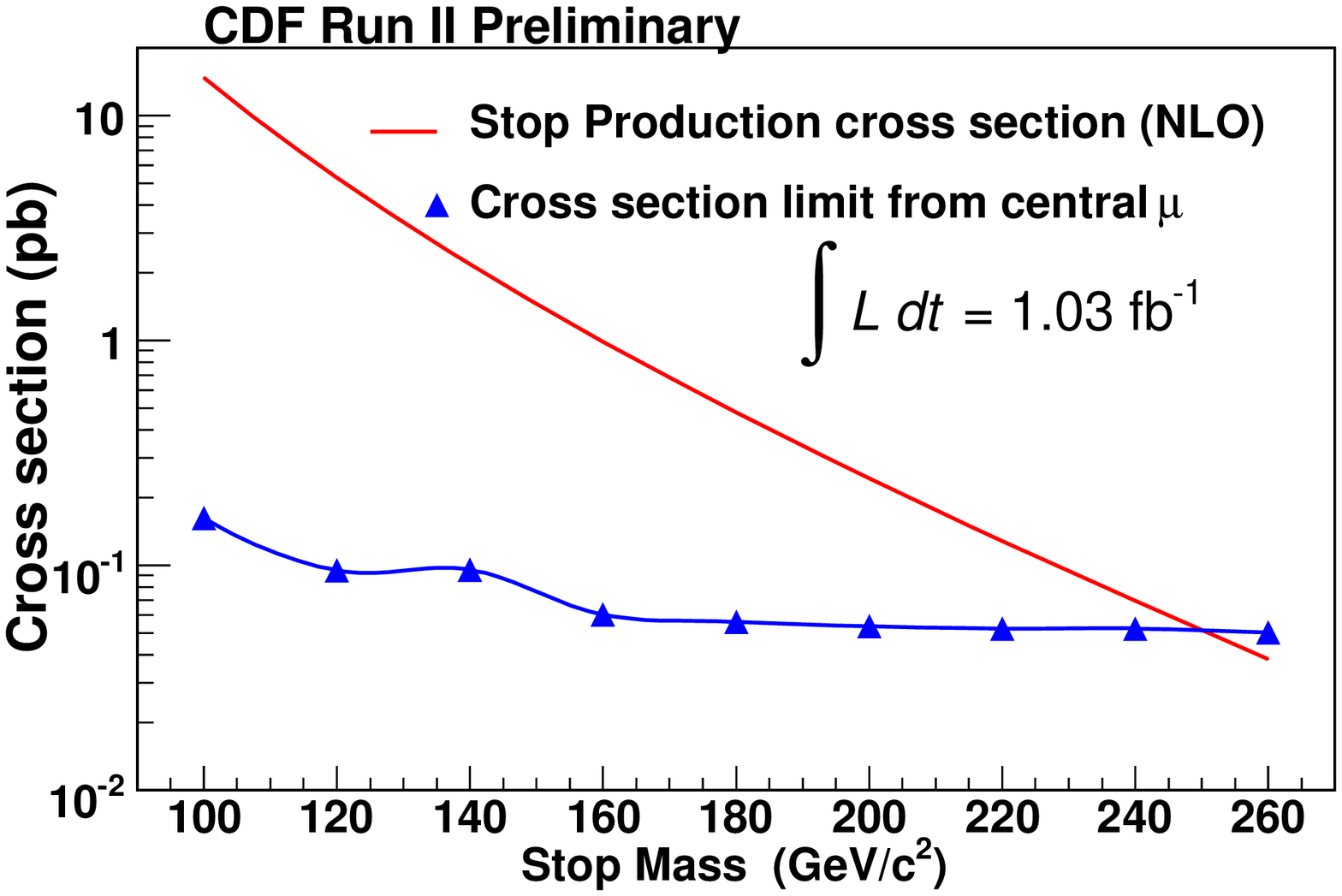}
\\
\parbox[t]{0.32\textwidth}{\caption{The 95\% C.L. exclusion region in the mass plane of
                                    $m(\schi^{0}_{1})$ vs $m(\sstop_{1})$ for the search
                                    of pair production of stop.}
\label{fig:stop}}
\hfill
\parbox[t]{0.32\textwidth}{\caption{The 95\% C.L. exclusion region in the $\schi^{0}_{1}$ lifetime
  vs $\schi^{0}_{1}$ mass plane for long lived $\schi^{0}_{1}$ search.}
\label{fig:delayphoton}}
\hfill
\parbox[t]{0.32\textwidth}{\caption{The 95\% C.L. upper limit stop production cross section as
function of the stop mass.}
\label{fig:champs}}
\end{figure}

Some particles predicted in SUSY carry electrical charge
and are massive (CHAMPs). These particles are expected to
be slow moving and are very penetrating, just like a ``slow muon''.
They could decay outside the detector is they are long lived.
CDF and D\O\ have searched for signatures of CHAMP particles
inside their detectors.
In the search the experiments look for events that contain
$\mu$-like particles that are slow moving.
CDF makes use of its Time-of-Flight (ToF) and tracking
detectors to measure the CHAMP candidates' velocities and momenta,
and then determine its masses.
D\O\ identifies the slow moving $\mu$-like particles based on
the timing measurements recorded in the muon detector.
No evidence of CHAMP is observed in the data of both experiments.
Within the SUSY model with one compactified extra dimension \cite{SUSYCED}
where the stop is the LSP, CDF excludes the mass of a stable stop
up to 250 GeV/c$^{2}$ (shown in Figure \ref{fig:champs}).
D\O\ presents its limits in two models.
In the GMSB model D\O\ assumes stau ($\tilde{\tau}$) is the CHAMP
particle, and they obtain an upper limit on the production cross section
from 0.62 pb to 0.06 pb for various $\tilde{\tau}$ mass.
In the stable chargino model (ref) where the chargino $\schi^{\pm}_{1}$
is assumed to be the CHAMP particle, D\O\ excludes the chargino
mass up to 174 GeV/c$^{2}$.

\section{Summary}
CDF and D\O\ have searched for SUSY in up to 1 fb$^{-1}$
of data sample, and have yet found evidence of its existence.
Some of the limits set on the SUSY model parameters are the
world's best (example the masses of the squarks and gluino).
The Tevatron machine and the experiments are now working optimally.
We expect more significant improvement in the SUSY searches
at the Tevatron with increase in integrated luminosity
and with smarter analysis techniques.

\section*{References}

\end{document}